\RequirePackage{fix-cm}
\documentclass{article}                     
\usepackage{appendix}
\usepackage{amsmath}
\usepackage{amssymb}
\usepackage{graphicx}
\usepackage{lineno}
\usepackage{array}
\usepackage{longtable}
\usepackage{natbib}
\usepackage[margin=1in]{geometry}
\usepackage{subcaption}
\usepackage{nicefrac}
\usepackage{makecell} 
\newcommand*\patchAmsMathEnvironmentForLineno[1]{%
\expandafter\let\csname old#1\expandafter\endcsname\csname #1\endcsname
\expandafter\let\csname oldend#1\expandafter\endcsname\csname end#1\endcsname
\renewenvironment{#1}%
{\linenomath\csname old#1\endcsname}%
{\csname oldend#1\endcsname\endlinenomath}}%
\newcommand*\patchBothAmsMathEnvironmentsForLineno[1]{%
\patchAmsMathEnvironmentForLineno{#1}%
\patchAmsMathEnvironmentForLineno{#1*}}%
\AtBeginDocument{%
\patchBothAmsMathEnvironmentsForLineno{equation}%
\patchBothAmsMathEnvironmentsForLineno{align}%
\patchBothAmsMathEnvironmentsForLineno{flalign}%
\patchBothAmsMathEnvironmentsForLineno{alignat}%
\patchBothAmsMathEnvironmentsForLineno{gather}%
\patchBothAmsMathEnvironmentsForLineno{multline}%
}

\usepackage{color}

\usepackage{ulem}

\usepackage{tikz}
\usepackage{pgfplots}
\usepgfplotslibrary{groupplots}
\usepackage{authblk}

\begin{document}

\title{Statistical investigations of flow structures in different regimes of the stable boundary layer
}


\author[1]{Nikki Vercauteren}
\author[1]{Vyacheslav Boyko}
\author[1]{Amandine Kaiser}
\author[2]{Danijel Belu{\v s}i{\'c}}

\affil[1]{FB Mathematik und Informatik, Freie Universit{\"a}t Berlin, 14195 Berlin, Germany}
\affil[2]{Swedish Meteorological and Hydrological Institute (SMHI), Norrköping, Sweden}



\maketitle

\begin{abstract}
A combination of methods originating from non-stationary timeseries analysis is applied to two datasets of near surface turbulence in order to gain insights on the non-stationary enhancement mechanism of intermittent turbulence in the stable atmospheric boundary layer (SBL). We identify regimes of SBL turbulence for which the range of timescales of turbulence and submeso motions, and hence their scale separation (or lack of separation) differs. Ubiquitous flow structures, or events, are extracted from the turbulence data in each flow regime. We relate flow regimes characterised by very stable stratification but different scales activity to a signature of flow structures thought to be submeso motions.



\end{abstract}
\pagebreak
\section{Introduction}
\label{intro}
The representation of the stable boundary layer (SBL) presents ongoing challenges, and modelling challenges increase with increasing stability \citep{Sandu:2013ia}. Among the more unknown situations are small wind speed scenarios in which the turbulence is weak and does not show significant dependence on the stratification. In such weak wind situations, turbulence typically becomes non-stationary and a spectrum of motions on the so-called sub-mesoscales is found to bridge the scale gap between the largest turbulent scales and mesoscales \citep{Anfossi:2005kn, Belusic:2010jc, Mahrt:2014wc}. Weak turbulence is found to be enhanced by these submeso motions \citep{Mahrt:2015es, Sun:2015km, Cava:2016ho}. Better understanding of the non-stationary enhancement mechanism is a necessary step towards improved SBL turbulence parameterisation.

Recent approaches focus on distinguishing flow regimes in which turbulence behaves differently. Based on observations, \cite{Sun:2012eo} identify a height dependent wind speed threshold that separates a regime in which turbulence increases slowly with increasing wind speed from a regime where turbulence increases rapidly with the wind speed. The weak turbulence, strongly stable regime is found to include cases where local shear-generated eddies are too small to interact with the ground and turbulence is not related to the bulk shear anymore. Theoretical findings also predict the appearance of two regimes based on the hypothesis that continuous turbulence requires the turbulence heat flux to balance the surface energy demand resulting from radiative cooling \citep{vandeWiel:2012ji, vandeWiel:2012jy, vandeWiel:2017ju}. A radiative heat loss that is stronger than the maximum turbulent heat flux that can be supported by the flow with a given wind profile will lead to the cessation of turbulence \citep{vandeWiel:2012ji}. This concept is used by \cite{vanHooijdonk:2015gh} to show that the shear over a layer of a certain thickness can predict SBL regimes when sufficient averaging of data is considered. 

The very stable regime is however more prone to be dominated by apparently random, sub-mesoscale wind accelerations that can generate local turbulence and lead to highly non-stationary flows \citep{Acevedo:2015iw}. Such local accelerations have been revealed by released fog elements and fine scale temperature measurements from fibre optic distributed temperature sensing \citep{Zeeman:2014ec}. Numerical studies have shown that finite perturbations imposed on the flow after cessation of turbulence can suffice to act as a regenerating mechanism for turbulence \citep{Donda:2015kh}. \cite{Donda:2015kh} further found a strong sensitivity of the turbulence recovery to the timing and amplitude of added perturbations, thereby motivating the need for better characterisation of sub-mesoscale motions and their effect on turbulence. Statistical analyses of the hydrodynamical equilibrium properties of the SBL flow revealed that the very stable regime is prone to long-term memory effects in the turbulence dynamics, suggesting a dynamically unstable flow \citep{Nevo:2017ir}. The long-term memory effects could be related to sub-mesoscale motions that can propagate for some time in very stable flow regimes due to weak turbulent mixing. Such memory properties in the turbulent observables suggest that very stable flow regimes need to be represented by high order closure models or stochastic processes \citep{Nevo:2017ir}. A statistical characterisation of sub-mesoscale flow structures and their transport properties would greatly help defining such a stochastic process.

Despite numerous case studies highlighting the local shear generation of turbulence due to wind speed accelerations connected to submeso motions \citep{Sun:2004tj, RomanCascon:2015ku, Mortarini:2017em}, the general understanding of non-turbulent motions on sub-mesoscales remains very limited. Analyses of the propagation direction of submeso motions revealed no tendency to follow the mean wind direction \citep{Lang:2017bu} and highlighted the difficulty to understand the origin of such features of the flow. To extend case studies to more general observations, \cite{Kang:2014do} developed a method to extract non-stationary motions from turbulent timeseries, regardless of the physical origin of the flow motions. Non-stationary flow structures from SBL data were subsequently categorised into three classes with similar characteristics \citep{Kang:2015hi}. The smoothest, wave-like structures were typically associated with stronger wind, active turbulence and weak stability. The two other classes associated with higher stratification were found to have predominantly sharp structures and one of them included step-like structures that were attributed to microfronts.

To further investigate the local shear generation of eddies, \cite{Vercauteren:2015fq} followed a data driven approach to identify regimes based on the relationship between turbulence and local wind variations on the sub-mesoscales. The regime identification was based on the Finite Element, Bounded Variation, Vector Autoregressive factor models (FEM-BV-VARX) clustering procedure \citep{Horenko:2010ce, OKane:2016fd}, which allows to explicitly consider external factors influencing the dynamics of the turbulent observables in the classification of flow regimes. The automatic procedure was developed to isolate periods in which turbulence is related to local acceleration of the flow due to propagating non-turbulent motions on the sub-mesoscales. Further analysis in one dataset revealed that one of two identified types of submeso-influenced regimes gathered cases in which a scale gap separated the smallest sub-mesoscales from the largest turbulence scale. In the second such regime, sub-mesoscales and turbulent scales seemed to overlap \citep{Vercauteren:2016kx}. 
Based on the classification of submeso-influenced flow regimes, the present study will characterise the statistics of submeso motions that occur in flow regimes characterised by different scale activity. The extraction of submeso motions will be based on the Turbulent Event Detection (TED) method proposed by \cite{Kang:2015hi}. The questions that will be addressed are the following: 
Is there a preferred type of submeso motion that interacts with turbulence? 
And does the frequency and type of submeso-motions change depending on the regimes of SBL turbulence?

\section{Data}
\label{sec:1}
Our study is based on sonic anemometers measurements from the Snow Horizontal 
Array Turbulence Study (SnoHATS, \cite{BouZeid:2010he}) and from the Fluxes 
over Snow Surfaces II (FLOSSII, \cite{Mahrt:2010gv}) datasets. The SnoHATS 
dataset was collected over a large flat glacier on top of a mountain range. The 
FLOSSII dataset was collected over a locally flat basin between two mountain 
ranges and includes several snow covered periods. Some measures of turbulence 
and sub-mesoscale activity are given for both sites in Table \ref{tab:sites}.
\subsection{SnoHATS}
\label{subsec:1.1}
\begin{figure}[h]
    \centering
    \begin{subfigure}[t]{0.45\textwidth}
        \includegraphics[width=\textwidth]{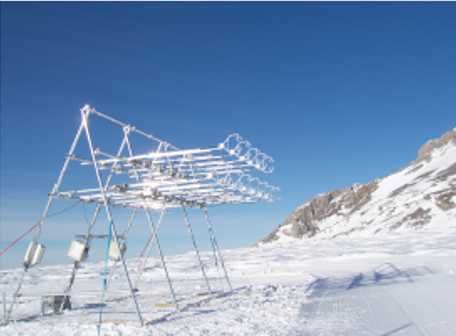}
    \end{subfigure}
    ~ 
    \begin{subfigure}[t]{0.44\textwidth}
        \includegraphics[width=\textwidth]{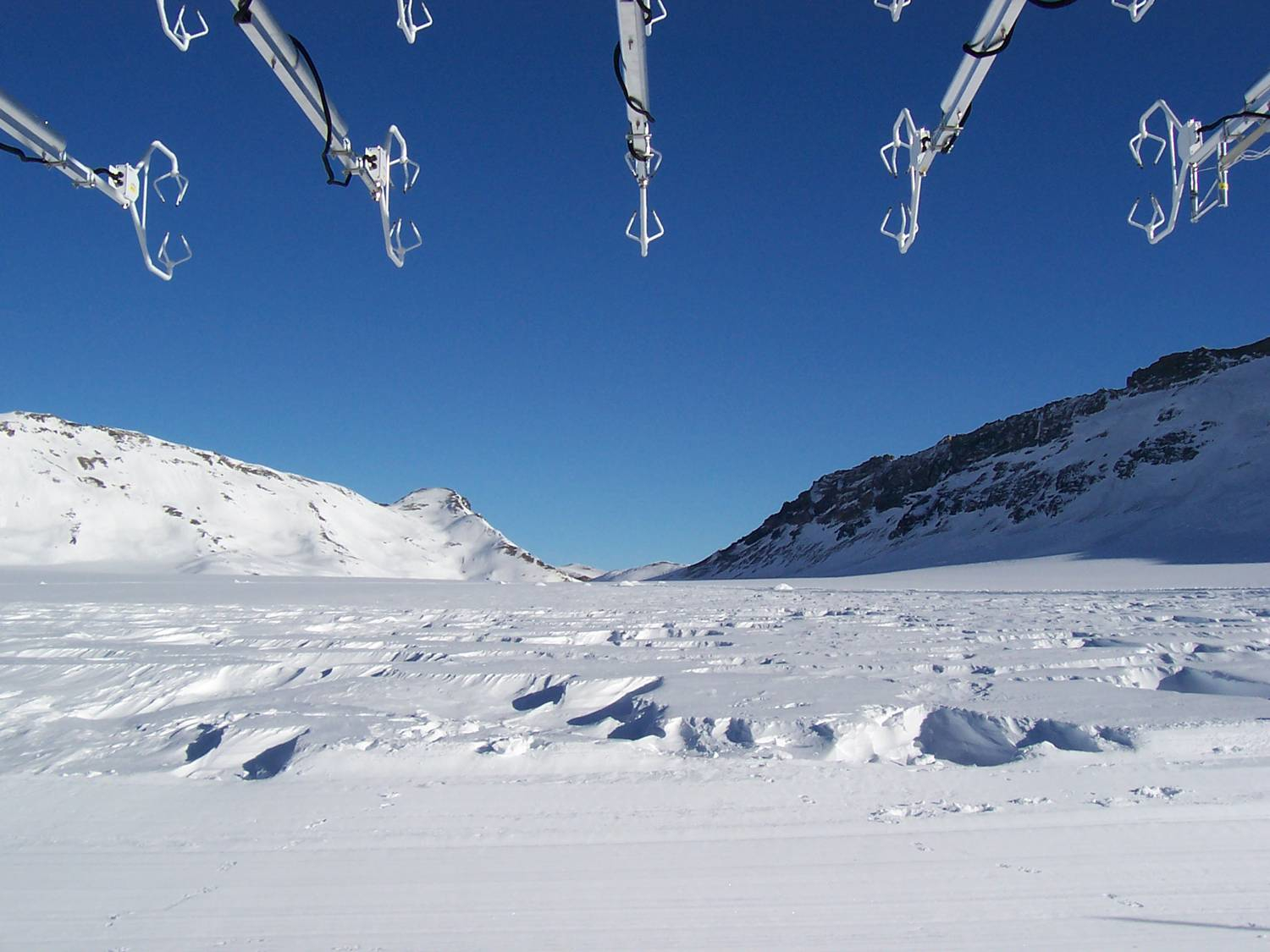}
    \end{subfigure}
    ~ 
    \caption{Setup of the SnoHATS field campaign. Left: Side view with the 12 instruments. Right: View in the direction of measurements showing the 1.5 km fetch. }\label{fig:SnoHATS_setUp}
\end{figure}

The data was collected over the Plaine Morte Glacier in the Swiss Alps from February to April 2006, at 2750m elevation (\citep{BouZeid:2010he}, data collected by the EFLUM laboratory at EPFL). The large flat glacier ensures long periods of stable stratification, and measurements were taken at a height varying between 2.82 m and 0.62 m, depending on snow accumulation. The setup, shown in Fig. \ref{fig:SnoHATS_setUp}, consists of  two vertically separated horizontal arrays of sonic anemometers, with a total of 12 sonic anemometers (Campbell Scientific, model CSAT3). The vertical separation between the upper and lower array is 77 cm (82 cm after March 17), while the horizontal separation between the instruments is 80 cm. The data analysis was restricted to wind directions within a $\pm$60š$^\circ$š angle relative to the streamwise sonic axis (corresponding to easterly winds), ensuring that data are not affected by the structure supporting the instruments. The resulting fetch consists of 1500m of flat snow. After removing data with unfavourable wind angles (outside the selected $\pm$60š$^\circ$ range) or low quality (snow-covered sonics, power outages, etc), about 15 non-continuous days of data remained available for the analysis. The 20Hz raw data were preprocessed and conditioned using axis rotations to correct for the yaw and pitch misalignments of the sonics, linear detrending and density correction. 
\subsection{FLOSSII}
\label{subsec:1.2}
The data was collected from 20 November 2002 to 4 April 2003 over a locally flat surface south of Walden, Colorado, USA, in the Arapaho National Wildlife Refuge. The surface consists of matted grass with brush upwind about 100 m. The grass was often covered by a thin snow layer during the field program. A tower collected measurements at 1, 2, 5, 10, 15, 20 and 30 m with Campbell CSAT3 sonic anemometers and the data from the second level (2 m) are used to identify flow regimes, extract and characterise events. The choice of the 2 m level is made to be similar to the SnoHATS data as well as to be enough above the ground to avoid dissipation of structures by small-scale turbulence near the surface. It also ensures that the data remain within the boundary layer, which can be very shallow in strongly stable conditions. Investigations of the
height dependence of flow regimes is left for future work. Here instead, the transport characteristics of different scales of motion will be analysed at different heights assuming that the regime affiliation is the same for all heights. The data set was quality controlled and segments of instrument problems and meteorologically impossible values were eliminated (Larry Mahrt, personal communication).
We restrict the analysis to night time data, taken between 18:00 and 7:00 (Local time). Flow regime identification based on the FEM-BV-VARX clustering methodology (see Section \ref{method_VARX}) ideally requires continuous data, however the dataset will consist of continuous night time data separated by gaps during the day. In order to maximise continuity of the dataset, nights with data gaps corresponding to more than 80 minutes (12 nights) as well as nights with wind flowing regularly through the measurement tower for periods longer than 5 minutes (51 nights) were removed from the analysis. The resulting 68 nights left for analysis have data gaps shorter than 1 minute and are deemed mostly uncontaminated. The short gaps are linearly interpolated. The 60Hz raw data is double rotated in to the mean wind direction based on 30  minutes average.  
%
%
\begin{table}
	\centering
	\caption{Site characteristics. Averaged values of the 30-min records for: 
	the standard deviation of the vertical velocity fluctuations  $\sigma_{w}$, 
	the wind speed $V$, the sub-mesoscale wind velocity $\hat{V}$ (defined 
	formally in Section \ref{method_VARX}), the percentage of the time where 
	the submeso-velocity scale is greater than the speed of the 30-min 
	averaged wind vector, and the submeso cross-stream velocity variance $\sigma^2_{vM}$ (defined formally in \cite{Vickers:2007hd}, equation (1)). The average include all instruments at the site.}
	\label{tab:sites} 
	\begin{tabular}{lcccccc}
		\hline\noalign{\smallskip}
		Site & $\sigma_{w}$ & $\sigma_{w}/V$ & $V$ & $\hat{V}$ & $\hat{V}/V>1 $ & $\sigma^2_{vM}$\\
		$ $ & $[m/s]$ & $[-]$ & $[m/s]$ & $[m/s]$ & $ [\%]$ & $[-]$ \\
		\noalign{\smallskip}\hline\noalign{\smallskip}
		SnoHATS & $0.18$ & $0.09$ & $2.68$ & $0.70$ & $4.6$ & $0.72$\\
		FLOSSII & $0.31$ & $0.06$ & $5.26$ & $0.74$ & $10.18$ & $0.35$ \\
		\noalign{\smallskip}\hline
	\end{tabular}
\end{table}

\section{Methods}
\label{sec:2}
Our analyses of flow structures in the SBL are based on two complementary methods. In a first step, flow regimes are identified based on the intensity of turbulent velocity fluctuations and their modulation by a sub-mesoscale wind velocity. The identification uses a data-clustering methodology based on a finite element, bounded variation, vector autoregressive factor method (FEM-BV-VARX) introduced by Horenko \citep{Horenko:2010ce, OKane:2013bt}. We hypothesise that the turbulence will sometimes be modulated by the wind variability on sub-mesoscales (typically in weak wind, strongly stable situations) and our goal is to automatically detect periods in which the sub-mesoscale wind velocity influences the turbulence \citep{Vercauteren:2015fq}. In the second step, we apply the Turbulent Event Detection method introduced by \cite{Kang:2014do, Kang:2015hi} to detect events in noisy timeseries. The type of turbulent event occurring will be analysed in each of the FEM-BV-VARX identified flow regime separately, thus giving indication on the type of submeso motions occurring in each of the flow regimes detected based on scale interaction properties.

\subsection{Classification of flow regimes}
\label{method_VARX}
In this section we briefly review the mathematical framework used to classify the flow regimes in terms of their scale interactions properties. For the full details of the mathematical framework, we refer to \cite{Horenko:2010ce}, while further details on its application to SBL flow regime classification can be found in \cite{Vercauteren:2015fq}. 

The FEM-BV-VARX method relates an observed variable of interest at a discrete time $t$ to its past history, and to the past history of external forcing variables. The classification of SBL flow regimes is based on the hypothesis that in some flow regimes, turbulence may be modulated to a large extent by sub-mesoscale motions. Those flow regimes are expected to correspond to weak wind, very stable periods. Our classification goal is to separate
cases during which the time evolution of turbulence is modulated by the time evolution of the sub-mesoscale
wind velocity from cases during which the response of turbulence to forcing by sub-mesoscales is different or
less apparent. More specifically, we assume that the evolution in time of the vertical velocity fluctuations $\sigma_w = \sqrt{\overline{w'w'}}$ (where the overbar denotes an averaging period of 1 minute and the prime denotes deviations from the average) can be approximated by several locally stationary statistical processes that are influenced by the sub-mesoscale horizontal wind velocity $\hat{V}$, defined on scales between 1 and 30 min. The sub-mesoscale mean wind speed is defined formally as 
\begin{equation}\label{Eq:Vsmeso}
\hat{V} = \sqrt{ \hat{u}^{2} + \hat{v}^{2} }\, , 
\end{equation}
where $\hat{\phi} = \overline{\phi} - [\phi]$, the overbar denotes a  1-min 
averaging time and the square brackets denote a 30-min averaging time, such 
that these fluctuations represent the deviations of the 1-min sub- record 
averages from the 30-min average.  The definition of sub-mesoscales is made 
because those are scales that typically correspond to non-turbulent motions in 
weak-wind SBL flows \citep{Mahrt:2012do}. Furthermore, the choice of the 1 min 
averaging time for the vertical velocity variance is a compromise between 
minimising loss of flux by larger-scale turbulent motions with windy conditions 
and minimising the contamination of the computed fluctuations by non-turbulent 
motions for weak-wind more stable conditions. \par 
The statistical processes representing the time evolution of $\sigma_w$ in the FEM-BV-VARX framework are vector autoregressive models with exogenous factors (VARX) and the sub-mesoscale wind velocity $\hat{V}$ is considered as the exogenous factor. Our analyses showed that models which included an autoregressive part did not give any reproducible solutions to the clustering problem. Hence we restrict our search to models including only the exogenous part:

\begin{equation} \label{eq:VARX}
\sigma_{w;t} = \mu (t) + B_{0} (t) \hat{V}_{t} + \cdots  + B_{m}(t) \hat{V}_{t-m \tau} + C(t) \epsilon_{t},
\end{equation}
where the process $\sigma_{w;t}$ is the time evolution of the 1-min vertical 
velocity variance measured at one location; the external factors $\hat{V}_{t}$ 
are the time evolution of the streamwise velocity on scales between 1 and 
30min. $\epsilon_{t} \colon [0,T] \to \mathbb{R}^{h} (h \ll n)$ is a noise 
process with zero expectation, the parameters $\mu$, $B$ and $C$ are time- 
dependent coefficients for the statistical process, $m$ is the memory depth of 
the external factor which needs to be estimated and $\tau$ is the discrete time increment (here one minute). The number of statistical 
 processes corresponds to the number of clusters; 
the assumption of local stationarity of the statistical process is enforced by 
setting a persistency parameter $C_{p}$, which defines the maximum allowed 
number of transitions between $K$ different statistical processes 
(corresponding to $K$ different values of the model coefficients in Eq. (\ref{eq:VARX})). The cluster 
states are indicated by a cluster affiliation function, which is calculated by 
the procedure. The assumption of local stationarity of the statistical process is equivalent to assuming that the dynamics consists of several persistent flow regimes. In other words, the characteristic fluctuation time scale of the data is assumed to be fast compare to the time scale at which switches between flow regimes occur. In each flow regime, an optimal process of the form given in Eq. (\ref{eq:VARX}) will provide a representation of the statistical modulation of the dynamics of the vertical velocity variance by the sub-mesoscale wind velocity. The reader is referred to \cite{Horenko:2010ce} for information 
regarding the minimisation procedure used to solve the clustering problem. More detailed explanations on the application of the classification scheme to SBL turbulence are given in \cite{Vercauteren:2015fq}. User 
defined parameters and their choice are discussed in Section 
\ref{subsec:results_varx}.

\subsection{Turbulent events detection}
\label{subsec:2.3}
The time series analysis methodology for turbulent event detection (TED) derived by \cite{Kang:2014do} aims at identifying non-stationary events or flow patterns in noisy time series. Instead of detecting signatures of known flow patterns in time series, the TED method detects flow structures as events that are significantly different from noise. In the context of time series resulting from turbulent quantities, the noise is taken as white and red noise. Indeed, statistical descriptions of turbulence as first suggested by \cite{Kolmogorov:1941vt} and \cite{Obukhov:1941wn} lead to the formulation of  stochastic models for the turbulent observables such that in the inertial subrange, Lagrangian velocities can be modelled by a Langevin equation (or Ornstein-Uhlenbeck process) with suitable drift and noise terms \citep{Thomson:1987ge}:
\begin{equation}
	\text{d}u = -\frac{u}{T}\text{d}t + \sqrt{C_0 \varepsilon}\, \text{d}W\,, \label{eq:Langevin_model}
\end{equation}
where $u$ is the velocity (or a turbulent observable), $T$ is the Lagrangian decorrelation time scale, $C_0$ is a universal constant and $\varepsilon$ is the the mean dissipation; $\text{d}W$ are increments of a Wiener process. As shown in \cite{Faranda:2014bd}, this model is in fact equivalent to an autoregressive process of order one (AR(1)) process (also known as red noise):
\begin{equation}
	u_t = \phi\, u_{t-1} + \psi_t\, , \label{eq:AR_model}
\end{equation}
where $t$ is a discrete time label, $\phi = 1 - \nicefrac{\Delta t}{T}$, and $\psi_t$ represents independent variables, normally distributed. 

In the SBL, gravity waves, transient drainage flows and other flow structures on sub-mesoscale will typically superimpose on the turbulence or affect its intensity, thereby inducing non-stationarity and hence departures from the idealised inertial subrange Langevin model (\ref{eq:Langevin_model}) or AR(1) model (\ref{eq:AR_model}). Deviation from AR(1) processes were in fact studied in \cite{Nevo:2017ir} to investigate the hydrodynamical equilibrium properties of turbulence in different SBL flow regimes, showing that intermittent or strongly stable regimes exhibit long memory effects in the turbulence dynamics. The core idea of the TED method is also to analyse deviations from AR(1) processes: in a first step, sequential subsequences of the time series $x(t)$ of turbulent observables are analysed using a sliding window of predefined length-scale $l$. The q-th subsequence is thus:
\begin{equation}
	x_q(t) = \left\lbrace x(t_q), \ldots x(t_{q+l-1})  \right\rbrace \, ,
\end{equation} 
where $1 \leq q \leq (n-l+1)$ and $n$ is the length of the time series $x(t)$. Events are defined as subsequences that are significantly different from white noise or from an AR(1) process. In practice, an AR(1) model is fitted to each detrended subsequence $x_q(t)$ and a test is performed on the model residuals to see whether they are uncorrelated. If this is not the case (i.e. if the residuals are not white noise), then $x_q(t)$ is defined as a potential event.
Additionally, non-stationary subsequences that exhibit a structural break are considered as potential events. Note that the noise process is not removed from the subsequence, meaning that the potential event consists of the raw subsequence. 

The TED approach assumes that the typical duration of an event is known, but its form is unknown. In the context of detecting submeso motions, this is appropriate since submeso motions can take many different forms that are poorly known, but the typical duration of events is on the scale of minutes to an hour. A complementary approach to detect events in noisy timeseries is to assume that the form of events is known, while the duration is unknown. This approach is proposed in \citep{Lilly:2017jx} where wavelet elements embedded in noise can detect isolated events of a known form. The reader is referred to  \cite{Kang:2014do}; \cite{Kang:2015hi} for full details on the TED method. Time scales considerations are discussed next.

\subsection{Averaging time within the TED approach}
\label{subsec:2.4}
In the TED method, the length of the time window $l$ has to be predefined. On the fastest scales, the signal is most likely purely turbulent, and block averaging on small enough scales accelerates calculations without loosing information on the sub-mesoscales. The choice of scale for the block averages of the turbulent observables will define the time increments of the AR(1) model in Eq. \ref{eq:AR_model}. Hence the averaging scale should be chosen such that the increments fall within the range of scales of inertial turbulence. As shown by the extended multiresolution decomposition (MRD) analyses in \cite{Vercauteren:2016kx}, scales faster than approximately $5-10$ seconds exhibit fluctuations characteristic of isotropic turbulence and block averaging within this time range represents an appropriate choice. 

In order to have results comparable to the analyses of \cite{Kang:2015hi}, we thus consider their choice of block averages of 6 seconds and a window length of 120 points (12 minutes). As discussed in \cite{Kang:2014do}, events can be detected on multiple overlapping windows, such that the maximal event length is not limited to one window length. The extended MRD results in \cite{Vercauteren:2016kx} highlight non-turbulent fluctuations in the range of 50 seconds to 20-30 minutes, depending on the flow regime. The window length of 12 minutes, with possibilities of longer events through overlapping windows, is hence deemed appropriate.
 
For the analysis of multiple scales, according to \cite{Kang:2014do}, keeping $l$ constant and block averaging the time series to a desired scale leads to better results. This is due to the fact that the test statistic applied for the white noise test depends on $l$ and keeping $l$ constant returns consistent results for all scales. Tests varying the size of block averages between 1 and 15 seconds (while keeping $l=120$ points for consistency) showed a large sensitivity of the event detection to the choice of scales, highlighting the difficulty of using automatic methods for analyses of submeso motions \citep{kaiser2016}. Block averaging on scales shorter than 3 seconds returned many short and insignificant events, while averaging on blocks longer than 9 seconds returned very few events. The range of scales between 4 and 8 seconds returned qualitatively consistent results. Acknowledging the sensitivity to the exact choice of an averaging time, we present our results using the aforementioned time scales - deemed physically appropriate by the MRD analyses - in the following section. Note that this averaging time scale is used only for the TED procedure, the FEM-BV-VARX clustering being based on 1-minute averages as explained earlier.

\section{Results}
\label{sec:results}

\subsection{Parameters selection for flow regime classification}
\label{subsec:results_varx}
The FEM-BV-VARX framework is used to classify flow regimes in the SnoHATS and in the FLOSSII datasets. The turbulence data under consideration in Eq. (\ref{eq:VARX}) is the time evolution of the 1-min vertical velocity variance measured at one location, $\sigma_w (t)$ ; the external factor is the time evolution of the streamwise velocity on scales between 1min and 30min at that location, $\hat{V}(t)$. For the SnoHATS dataset, this location corresponds to one of the 12 sonic anemometers (results showed little sensitivity to the choice of instrument). For the FLOSSII dataset, the location is chosen as the second level (2 m), corresponding to a height similar to the SnoHATS data. 
 
User defined parameters include the maximum memory depth $m$ for the forcing variable, the number of clusters $K$ and the persistency parameter $C_{p}$, which limits the number of transitions between the clusters. The maximum memory lag that we use in this model is determined by a priori calculation of the partial autocorrelation function (pacf) for the variables $\sigma_w$ and $\hat{V}$ \citep{Brockwell:2002ur}. The correlation between the time series drops on average after a few minutes, and was set to $m = 3$ for the SnoHATS dataset and $m=6$ in the FLOSSII dataset (based on the average pacf over 68 nights). To determine the optimum number of $K$ and $C_{p}$, multiple models are fitted for varied values of the parameters $K$ and $C_{p}$. The Akaike Information Criteria (AIC, see \cite{Brockwell:2002ur}) was used in \cite{Vercauteren:2015fq} to select the optimal number of cluster states $K= 4$ and the persistency parameter $C_{p}=20$ for the SnoHATS data. 

For the FLOSSII dataset however, the AIC exhibits asymptotic behaviour towards zero for all models in the investigated parameter space ($K=2,3,4,5,7$ and $C_{p}=[2,302]$) and cannot be used as a selection criteria. Instead, the optimal model parameters are selected as those that minimise the correlation between the signal $\sigma_w$ and the model residuals $\epsilon_{t}$, while maximising the amount of variance of the signal explained by the model. Increasing the parameters beyond $K=3$ and $C_{p}=150$ does not reduce the correlation in the residuals and does not increase the modelled variance. Thus the choice of $K=3$ and $C_{p}=150$ is considered as an optimal model. The amount of variance of $\sigma_{w}(t)$ explained by the VARX model in the three clusters is 0.8\%, 3\% and 9.5\%.

Analysis of the model residuals in the three clusters however showed that the error distribution in the cluster corresponding to the largest explained variance of 9.5\% was not Gaussian. This cluster gathers the smallest values of $\sigma_w$ and has the most interaction between sub-mesoscales and vertical velocity fluctuations and we want to classify the dynamical interactions more accurately. Therefore, we select the time series in this specific cluster and classify it with the FEM-BV-VARX methodology further into two distinct clusters. This strategy leads to error distributions that are closer to normally distributed in the two subsequent clusters. Selecting only those periods of larger dynamical interactions between $\sigma_{w}$ and $\hat{V}$ enables a second level clustering which differentiates the dynamical interactions between $\sigma_{w}$ and $\hat{V}$ and not just the average intensity of the turbulent fluctuations. \par
The different performance of the AIC as a selection criterion can be explained by the different characteristics of both datasets in relation to the statistical model assumption. The VARX models to be fitted consist by assumption of an average value of $\sigma_{w}$ ($\mu (t)$ in Eq. \ref{eq:VARX}) and of the past history of $\hat{V}$. If the past history of $\hat{V}$ is unrelated to the dynamics of $\sigma_{w}$, as could be expected from weakly stable periods, the clustering of data will result mainly from classification of the mean value of $\sigma_{w}$. In this case of classification based on the mean value of $\sigma_w$, choosing the optimum based on standard information criteria resulted in an overfitted solution. The SnoHATS dataset includes a large number of strongly stable periods during which the modulation of $\sigma_w$ by $\hat{V}$ is significant (see table (\ref{tab:causality}) and next paragraph), thus the statistical modelling assumption is appropriate, and selecting the optimal clustering result based on standard information criteria is sufficient. During FLOSSII, strongly stable periods were observed but strongly outnumbered by periods of stronger winds or cloud cover. In those weakly stable periods, the dynamics of $\sigma_w$ is barely related to that of $\hat{V}$. The two step procedure followed here enables first to isolate periods of low mean values of $\sigma_w$, and then to classify those periods according to the dynamical modulation of the signal by $\hat{V}$. The decision of classifying the FLOSSII data exhibiting modulation by $\hat{V}$ into two clusters (and not more) is arguably subjective, but the decision of a second level classification is based on analysis of the noise and thus on an objective criterion. We decided not to extend the second level of clustering to more flow regimes, partly for simplicity. Further analyses will highlight differences in the obtained flow regimes and the classification is deemed informative. Note that each dataset is thus classified in four distinct regimes, which do not necessarily correspond to the same flow types at both sites. In the rest of the paper, the flow regimes obtained at SnoHATS will be denoted by S1-S4, and at FLOSSII by F1-F4. The regimes are numbered following increasing median stability of the flow.

The model coefficients fitted in each cluster give a quantitative indication of the scale interactions at both sites in each flow regime. The VARX model in Eq. (\ref{eq:VARX}) contains a total of 6 parameters for SnoHATS, and 9 parameters for FLOSSII, where the difference is due to the different maximal memory depth $m$ set from the analysis of the pacf: $\mu$ corresponds to the mean value of $\sigma_w$, $B_0$ to $B_3$ (for SnoHATS) or $B_0$ to $B_6$ (for FLOSSII) are the weights associated with the past history of the external factor $\hat{V}$ and $C$ is the weight associated to the noise part of the model. In order to compare the relative weight of the mean versus the external factor in each statistical model, we normalise each parameter by the mean value $\mu$ of the corresponding model. We then compute the norm $||\textbf{B}_{Si}/\mu_{Si}||$ (resp. $||\textbf{B}_{Fi}/\mu_{Fi}||$ for FLOSSII) of the vector $\left( B^{Si}_0/\mu_{Si}, \cdots, B^{Si}_3/\mu_{Si} \right)$ (resp. $\left( B^{Fi}_0/\mu_{Fi}, \cdots, B^{Fi}_6/\mu_{Fi} \right)$ for FLOSSII), where $\mu_{Si}$ and $\textbf{B}_{Si}$ are associated to the model coefficients in cluster $Si$, to estimate the relative weight of the external forcing compared to the mean in each statistical model. Note that with this normalisation, the weight of the mean is always 1. The values obtained for each flow regime at both sites are listed in table (\ref{tab:causality}). The increasing values for increasing stability denote that the more stable cases show more statistical causality between sub-mesoscales of motion and turbulence. The method thus captures subtle differences in scale interactions between different regimes. The results also highlight that the data at FLOSSII are characterised by less modulation of the turbulence by the sub-mesoscale wind velocity than the data at SnoHATS, reflecting the difficulty to obtain an optimal number of clusters at FLOSSII based on our scale interaction hypothesis.

Flow characteristics are given for each cluster in Table \ref{tab:clusters}. The gradient Richardson number 
\begin{equation}\label{Eq:Ri}
Ri = (g/ \Theta_{0}) \frac{\partial \bar{\theta} / \partial z }{\left(\partial \bar{\bold{V}} / \partial z \right)^2 },
\end{equation}
is used for indicative assessment of the stability properties in each regime. In (\ref{Eq:Ri}), $g$ is the gravity acceleration, $\theta$ is the potential temperature ($\Theta_{0}$ being the averaged one over the record), $\bold{V}$ is the wind vector, and the overline denotes a time average of one minute. The vertical gradients are calculated using the averages of the upper and lower sensors for the SnoHATS data, and using the 1m and 10m levels for the FLOSSII data. The median and quartiles of $Ri$ in each cluster (Table \ref{tab:clusters}) shows that weakly stable periods are separated from strongly stable periods by the FEM-BV-VARX procedure, with however large overlaps in the distributions of $Ri$ in the different clusters. Since the classification is based on the modulation of the turbulence by the sub-mesoscale wind velocity, this separation (along with the statistical causality results in table (\ref{tab:causality})) strengthens the hypothesis that modulation of the turbulence by submeso motions differs between the weakly and very stable regimes. The overlap of distributions of $Ri$ and their significant spread highlights the difficulty of defining a threshold based on $Ri$ for distinguishing flow regimes. Especially in the most stable flow regimes, $Ri$ values might be highly variable in time and may not be the right indicator of the turbulence level. Indeed, in a context of strong modulation by a constantly changing sub-mesoscale wind velocity, the turbulence will likely not be in statistical equilibrium with the wind forcing, and thus the $Ri$ number does not alone suffice to characterise the turbulence. Analyses of the anisotropy characteristics of turbulence in the FLOSSII dataset give further indication of a lack of statistical equilibrium of the turbulence in flow regimes F3 and F4 \citep{Vercauteren:2019qj}.

\begin{table}
	\centering
	\caption{Statistical causality between the signal $\sigma_w$ and the external factor $\hat{V}$ as quantified by the relative weights of the model coefficients in Eq. (\ref{eq:VARX}), in each cluster. The norm of the vector of weights associated to the sub-mesoscale wind velocity $\left(B_0, \cdots B_m\right)$ is normalised by the weight of the mean part in the model, $\mu$, for each individual model or cluster.}
	\label{tab:causality} 
	\begin{tabular}{lccc}
		\hline\noalign{\smallskip}
		
		Site & Flow regime &  $||\textbf{B}_{Si}/\mu_{Si}||$ or $||\textbf{B}_{Fi}/\mu_{Fi}||$   \\
		\noalign{\smallskip}\hline\noalign{\smallskip}
		SnoHATS & S1   & $0.11$  \\
		SnoHATS & S2  & $0.22$ \\
		SnoHATS & S3  & $0.92$ \\
		SnoHATS & S4  & $2.20$ \\
		\hline\noalign{\smallskip}
		FLOSSII & F1 &  $0.04$   \\
		FLOSSII & F2 &  $0.04$   \\
		FLOSSII & F3 &  $0.11$   \\
		FLOSSII & F4 &  $0.29$  \\
		\noalign{\smallskip}\hline
	\end{tabular}
\end{table}

\begin{table}
	\centering
	\caption{Flow characteristics in each flow regime. Richardson number, wind speed, sub-mesoscale wind speed (Median (first and third quartiles) given for each value), and the percentage of times where the submeso wind velocity exceeds the wind velocity. All values are based on one-minute averaged data.}
	\label{tab:clusters} 
	\begin{tabular}{lccccc}
		\hline\noalign{\smallskip}
		Site & Flow regime & $Ri$  & $V (m.s^{-1})$ & $\hat{V} (m.s^{-1})$ & $\%\frac{\hat{V}}{V}>1 $  \\
		\noalign{\smallskip}\hline\noalign{\smallskip}
		SnoHATS & S1 & $0.02 \ (0.01, 0.05)$ & $5.6 \ (3.9, 7.1)$ & $ 1.6 \ (1.0, 2.4)$  & $7.8$ \\
		SnoHATS & S2 & $0.12 \ (0.06, 0.23)$ & $4.2 \ (3.2, 5.2)$ & $ 0.8 \ (0.5, 1.2)$ & $4.8$\\
		SnoHATS & S3 & $0.29 \ (0.10, 0.80)$ & $2.1 \ (1.3, 3.0)$ & $ 0.8 \ (0.5, 1.3)$ & $12.8$\\
		SnoHATS & S4 & $0.67 \ (0.28, 1.76)$ & $1.9 \ (1.1, 2.7)$ & $ 0.4 \ (0.3, 0.7)$ & $6.5$\\
		\hline\noalign{\smallskip}
		FLOSSII & F1 & $0.03 \ (0.02, 0.04)$ & $9.0 \ (7.8, 10.3)$ & $0.59 \ (0.27, 1.07)$ & $0.19$  \\
		FLOSSII & F2 & $0.07 \ (0.05, 0.10)$ & $5.7 \ (4.8, 6.7)$ & $0.43 \ (0.20, 0.79)$ & $0.31$  \\
		FLOSSII & F3 & $0.14 \ (0.10, 0.24)$ & $3.6 \ (2.7, 4.4)$ & $0.46 \ (0.20, 0.90)$ & $5.45$  \\
		FLOSSII & F4 & $0.58 \ (0.26, 1.55)$ & $1.3 \ (0.7, 2.1)$ & $0.40 \ (0.18, 0.76)$ & $16.63$  \\
		\noalign{\smallskip}\hline
	\end{tabular}
\end{table}

\subsection{Scale interaction properties}
\label{subsec:mrd}
\begin{figure} 
\centering
\includegraphics[width=\textwidth]{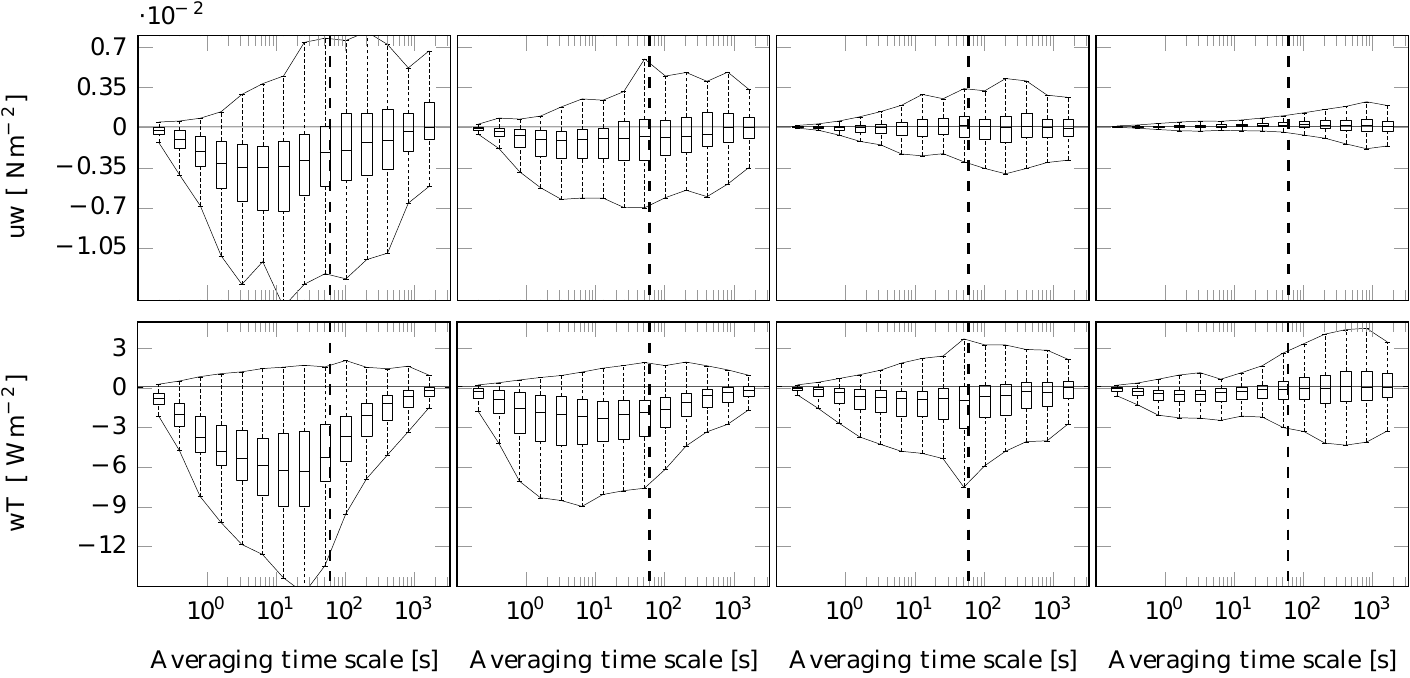}
\caption{MRD cospectra of the SnoHATS dataset. Top panels: momentum flux cospectra, bottom panels: heat flux cospectra. From left to right: Regime S1-S4. The major dashed vertical lines in every 
panel mark the one minute scale. Each panel contains box plots representing the distribution of the MRD flux on a corresponding scale. The boxes represent the 25th and 75th percentiles, and the whiskers 
across the scales are connected with a solid line. The horizontal line in each 
box shows the median. The statistics are calculated based on all 30-min periods within a flow regime. From S1 to S4, there are respectively 111, 148, 111 and 343 individual periods.}
\label{Fig:MRD_S}       
\end{figure}

\begin{figure} 
\centering
\includegraphics[width=\textwidth]{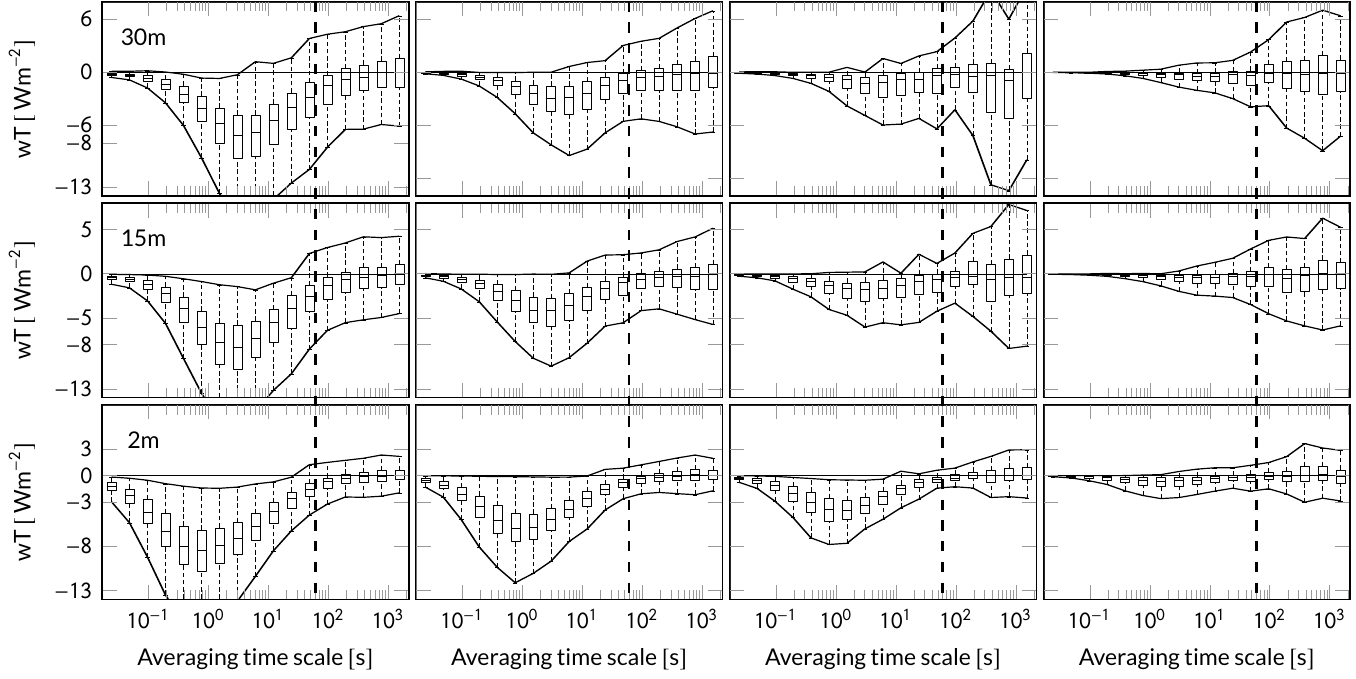}
\caption{MRD heat flux cospectra of the FLOSSII dataset. The top, middle and bottom rows
correspond respectively to the measurement heights of 30 meters, 15 meters and two meter. From left to right: regime F1 to F4, classified based on the two meter height measurements. The major dashed vertical lines in every 
panel mark the one minute scale. Each panel contains box plots representing the distribution of the wT on a corresponding scale. The boxes represent the 25th and 75th percentiles, and the whiskers 
across the scales are connected with a solid line. The horizontal line in each 
box shows the median. The statistics are calculated based on all 30-min periods within a flow regime, at the given height. From F1 to F4, there are respectively 401, 489, 86 and 215 individual periods.}

\label{Fig:MRD_F_wT}       
\end{figure}

\begin{figure}
\centering 
\includegraphics[width=\textwidth]{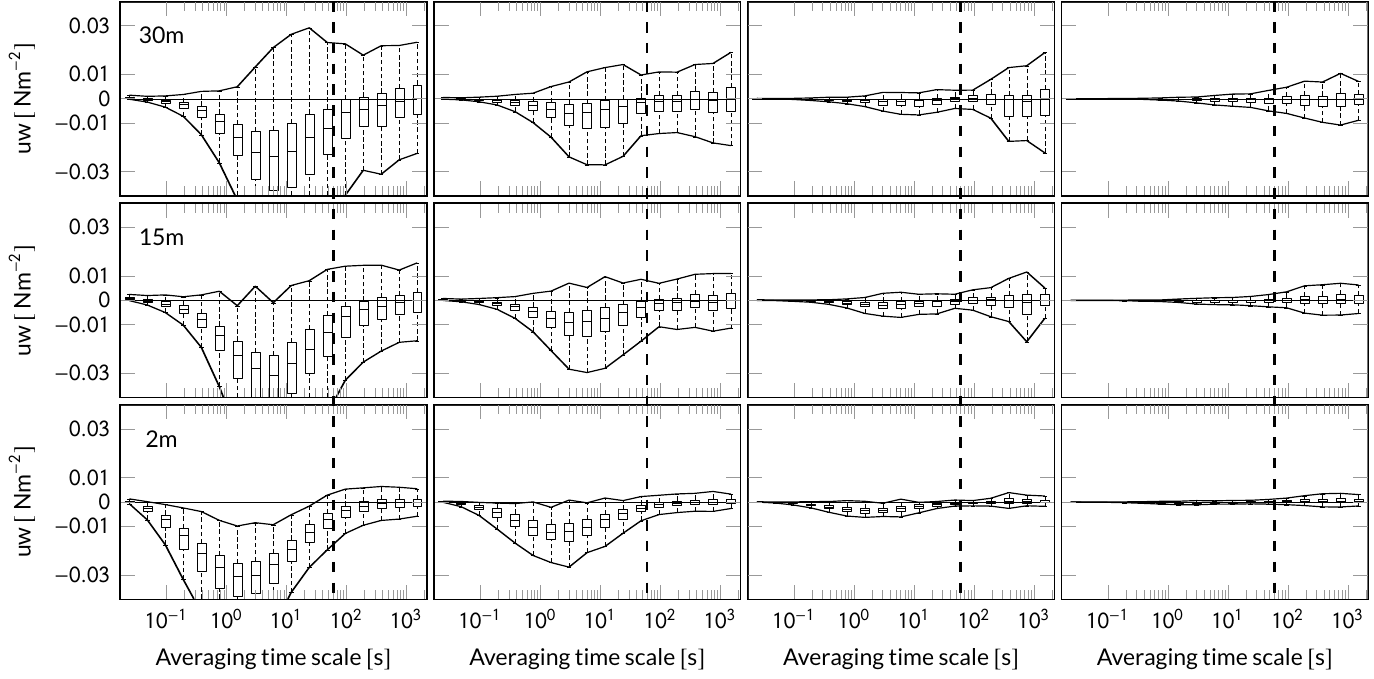}
\caption{MRD momentum flux cospectra of the FLOSSII dataset. The top, middle and bottom rows
correspond respectively to the measurement heights of 30 meters, 15 meters and two meter. From left to right: regime F1 to F4, classified based on the two meter height measurements. The major dashed vertical lines in every 
panel mark the one minute scale. Each panel contains box plots representing the distribution of the uw on a corresponding scale. The boxes represent the 25th and 75th percentiles, and the whiskers 
across the scales are connected with a solid line. The horizontal line in each 
box shows the median. The statistics are calculated based on all 30-min periods within a flow regime, at the given height. From F1 to F4, there are respectively 401, 489, 86 and 215 individual periods.}

\label{Fig:MRD_F_uw}       
\end{figure}

In each identified regime of near-surface SBL turbulence, the transport properties of different scales of motion are assessed using a multiresolution flux decomposition (MRD) \citep{Vickers:2003tw}. The scales activity is shown in Fig. \ref{Fig:MRD_S}-\ref{Fig:MRD_F_uw}, based on MRD analyses of the heat flux and momentum flux. In each figure, boxplots show the median and quartiles of MRD fluxes calculated based on all sampled 30-min windows in each flow regime. The MRD cospectra of the SnoHATS data are shown in Fig. \ref{Fig:MRD_S} for the flow regimes S1 to S4. The 
MRD cospectra of the FLOSSII data are shown for the heights 
of 2 m, 15 m and 30 m in Fig. \ref{Fig:MRD_F_wT} (for the sensible heat fluxes) and Fig. 
\ref{Fig:MRD_F_uw} (for the momentum fluxes) for the four classified flow regimes F1-F4 
(from left to right). The dashed vertical line shows the one minute 
average scale as a reference. 
In all flow regimes, the median MRD show an increased negative contribution with increasing scales until a maximum, followed by a decrease and finally crossing the zero flux line. This is the signature expected from the turbulent contribution and the scale at which the MRD heat flux first reaches zero is typically used to estimate the averaging scale required to sample the turbulent flux. 

The MRD of the heat fluxes show that while the averaged impact of a multitude of sub-mesoscale contributions to the heat fluxes are very small, individual sub-mesoscale contributions can be more important to the overall heat flux during a selected 30-min time period than turbulent contributions. This is apparent from the variability of the sub-mesoscale range of the MRD, visualised by the interquartile region of the figures at the sub-mesoscales. As the stability is increasing going from regime S1 to S4 (Fig. \ref{Fig:MRD_S}) or F1 to F4 (Fig. \ref{Fig:MRD_F_wT} and \ref{Fig:MRD_F_uw}), the magnitude of the transport by the turbulent scales reduces, while the variability of the transport by sub-mesoscales (i.e. the interquartile range at those scales) increases. As such, the dynamics becomes more influenced by the sub-mesoscales as stability increases. As the turbulence is collapsing, a state is reached where the magnitude of the heat flux due to sustained turbulent scales is overpassed by the local activity of individual submeso motions (Regime S4 and Regime F4). In the most stable regimes the local activity of individual submeso motion can be greater than that of the turbulent scales. In individual windows, sub-mesoscales thereby can represent the dominant contribution to the heat transport, although those scales do not systematically contribute to the mean fluxes.

Deeper analyses of the activity of different scales of motions in the SnoHATS flow regimes were presented in  \cite{Vercauteren:2016kx}. Extended MRD analyses suggested a likely direct transfer of energy from the sub-mesoscale horizontal velocity fluctuations to turbulent vertical velocity fluctuations in regime S3 and S4. Moreover, the analyses suggested that a scale gap separates sub-mesoscale motions from turbulence in S4, whereas flux variability was found to be more continuous in scale in S3 without a scale gap. 

With flow regimes classified according to the interactions between submeso activity and turbulent vertical velocity fluctuations in both datasets, the next section presents the characteristics of submeso motions identified by the TED method in each of the classified flow regimes.    

\subsection{Characteristics of events in different flow regimes}
\label{subsec:FEM_BV_VARX-regime}

Using the window size $l=120$ and $6 s$ averaged data, the first step of the TED method yields 1793 events in the SnoHATS temperature time series obtained by all 12 sonic anemometers, and 702 events in the FLOSSII 2-m temperature time series. A detailed analysis of the events extracted from the 2-m temperature time series of the FLOSSII data is given in \cite{Kang:2015hi}, presenting an event clustering approach based on event features which is not repeated here. The physical characteristics of turbulent events will be presented here for each of the FEM-BV-VARX flow regimes S1-S4 and F1-F4. 

\begin{figure}[!h]
	\centering
	\begin{subfigure}[b]{0.49\textwidth}
		\includegraphics[width=\textwidth]{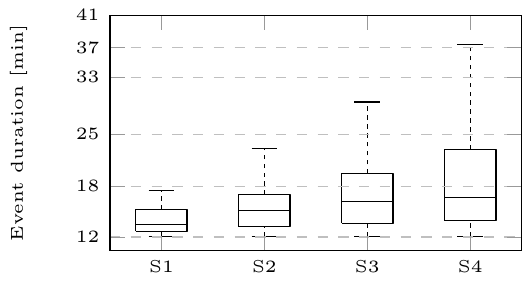}
		\caption{SnoHATS.}
		\label{sfig:eventDuration_SnowHATS}
	\end{subfigure}
	\begin{subfigure}[b]{0.49\textwidth}
		\includegraphics[width=\textwidth]{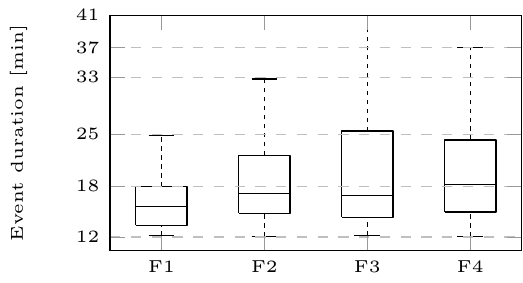}
		\caption{FLOSSII.}
		\label{sfig:eventDuration_FLOSSII}
	\end{subfigure}
	\caption{Boxplots of the events duration shown for each FEM-BV-VARX regime 
	separately and for each site.}
	\label{fig:eventDuration}       
\end{figure}

The time series within each regime are discontinuous and the TED method is applied to all continuous portions of the time series individually. Based on a $Ri$ number classification, \cite{Kang:2015hi} found that events occurred with similar frequencies for different stability ranges in the FLOSSII dataset. When comparing the frequency of occurrence of events in the flow regimes classified according to their scale interaction properties however, differences appear both in the SnoHATS data and in the FLOSSII data (Table \ref{tab:freqTED}). In the SnoHATS data, the events account for less than 10 $\%$ of the total time in the two regimes identified as little influenced by submeso motions and weakly stable (Regime S1: 7.5$\%$ and Regime S1: 9.2$\%$), whereas events account for 14.2$\%$ of the total time in Regime S3 and 20.4$\%$ of the total time in Regime S4. These two regimes are characterised by a median value of $Ri$ larger than $0.25$ (see Table \ref{tab:clusters}). Note that despite the high percentage of events in S4, the percentage of cases where the sub-mesoscale wind velocity is higher than wind speed is smaller in S4 than in S3 (Table \ref{tab:clusters}). Similarly in FLOSSII, the most turbulent regime F1 exhibits the lowest frequency of events (13.7 $\%$), while the most stable regimes (Regime F3 and F4) have the highest frequency of events (above 38 $\%$). A significant difference is however found in Regime F2 (also weakly stable) in which the frequency of events is large (35.0 $\%$). It could be that submeso motions are well represented in this regime, but that the mean shear is strong enough for the turbulence not to be affected by the sub-mesoscale wind fluctuations. In fact \cite{Vickers:2007hd} showed that the cross-wind velocity variance due to sub-mesoscale motions systematically increased with increasing wind speed at the FLOSSII site. This was speculatively attributed to enhanced generation of topographically induced motions by a nearby ridge, and could partly explain the higher percentage of events for the higher wind speed regime F3, when compared to the more flat terrain features of SnoHATS. Figure \ref{fig:eventDuration} further shows the event duration in the four flow regimes in SnoHATS and FLOSSII. The event duration increases with increasing regime affiliation number (i.e. with increasing stability and scale interactions) at SnoHATS, and is not very variable at FLOSSII expect for shorter events in the most turbulent regime F1. 
\begin{table}[!h]
	\centering
	\caption{Frequency of occurrence of the events for the regimes S1-S4 of SnoHATS and F1-F4 of FLOSSII. The numbers denote the percentage of the time detected as event within the total time of the corresponding regime.}
	\label{tab:freqTED} 
	\begin{tabular}{ccccc}
		& \multicolumn{4}{c}{SnoHATS/FLOSSII}\\
		\hline\noalign{\smallskip}
		& S1/F1 & S2/F2 & S3/F3 & S4/F4 \\
		\hline\noalign{\smallskip} 
		events [$\%$]  & $7.5/13.7$ & $9.2/35.0$ & $14.2/43.4$ & $20.4/38.2$\\
		\hline\noalign{\smallskip}
	\end{tabular} 
\end{table}

\begin{figure}[!th]
    \centering
    \begin{subfigure}[b]{0.45\textwidth}
    	\centering
        \includegraphics[width=\textwidth]{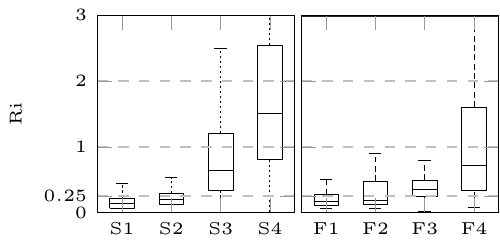}
        \caption{}      
        \label{sfig:Physical_chara_Rib}      
    \end{subfigure}
    \begin{subfigure}[b]{0.45\textwidth}
        \includegraphics[width=\textwidth]{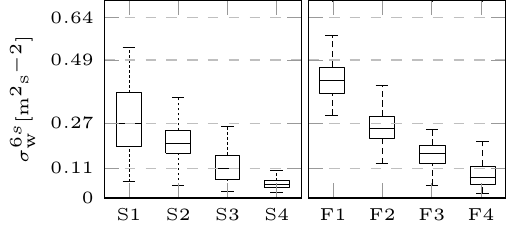}
           \caption{}    
        \label{sfig:Physical_chara_ww}
    \end{subfigure}
    ~ 
    \begin{subfigure}[b]{0.45\textwidth}
        \includegraphics[width=\textwidth]{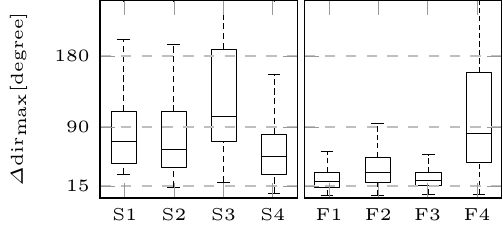}
           \caption{}    
        \label{sfig:Physical_chara_Dir}
    \end{subfigure}
    \begin{subfigure}[b]{0.45\textwidth}
        \includegraphics[width=\textwidth]{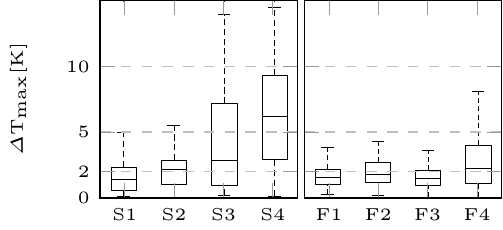}
           \caption{}    
        \label{sfig:Physical_chara_T}
    \end{subfigure}
    ~ 
    \begin{subfigure}[b]{0.45\textwidth}
       	\includegraphics[width=\textwidth]{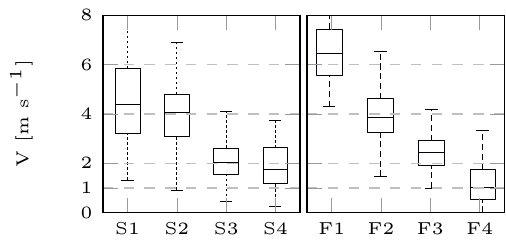}
	   \caption{}    
       	\label{sfig:Physical_chara_V}
    \end{subfigure}
    \begin{subfigure}[b]{0.45\textwidth}
       	\includegraphics[width=\textwidth]{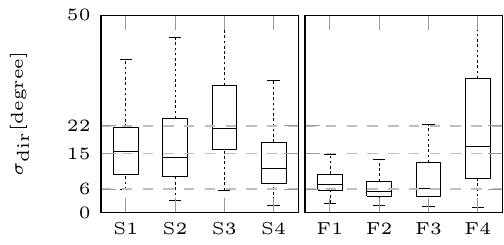}
	   \caption{}    
       	\label{sfig:Physical_chara_Wst}
    \end{subfigure}
    ~ 
    \begin{subfigure}[b]{0.45\textwidth}
       	\includegraphics[width=\textwidth]{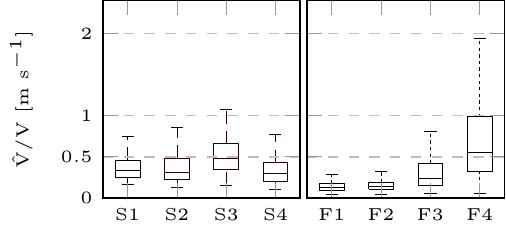}
	   \caption{}    
       	\label{sfig:Physical_chara_Era}
    \end{subfigure}
    \begin{subfigure}[b]{0.45\textwidth}
       	\includegraphics[width=\textwidth]{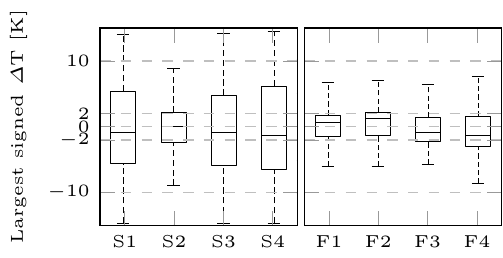}
	   \caption{}    
       	\label{sfig:Physical_chara_dir}
    \end{subfigure}
    \caption{ Boxplots of the TED events physical properties in each flow regime for SnoHATS (S1-S4) and FLOSSII (F1-F4). The line in each box represents the median over all events in the corresponding flow regime, while the bottom and top of the box are the 25th and 75th percentiles. (a) Richardson number, (b) Vertical velocity variance, (c) Largest wind direction change during the event, (d) Largest temperature change during the event, (e) Mean wind speed, (f) Standard deviation of the wind direction, (g) Sub-mesoscale to mean wind ratio, (h) largest signed temperature change during the event. In panels (a), (b), (e), (f) and (g), the values for individual TED events are calculated as the mean over the duration of the event, and a 6s averaging interval is used for variance and standard deviation calculations. In panels (c), (d) and (h), the largest change refers to the maximum difference between two consecutive 6s intervals during the event.}\label{fig:Physical_chara}
\end{figure}

The main physical characteristics of the events found in the different flow regimes are shown in Figure \ref{fig:Physical_chara}. The statistics of all events occurring in each flow regime are shown for the event averaged values of $Ri$, of the wind speed, of $\sigma_w$ (where the variance is calculated based on 6s intervals), and of the ratio of sub-mesoscale wind velocity to wind velocity. Note that the statistics differ from those given for the flow regimes in Table \ref{tab:clusters}, as they correspond only to TED event periods within a flow regime in Figure \ref{fig:Physical_chara}. Also shown is the standard deviation of the wind direction during events (based on 6s intervals to compute the wind direction), and the largest change in temperature and in wind direction during the event (defined as the difference between the maximum and minimum values during the event), so as to detect the signature of microfronts or sharp changes in wind direction. As expected since Regime S1, S2, F1 and F2 correspond to weakly stable periods with little influence of sub-mesoscales on the turbulence, the Richardson numbers during events are small, the wind speed is relatively high and the vertical velocity fluctuations are large in those regimes. The vertical velocity fluctuations decrease with increasing regime affiliation number, corresponding to increasing stability and modulation by sub-mesoscale wind velocity. Events in Regimes S4 and F4 are associated with very little turbulent vertical velocity fluctuations.

The largest differences appear in the behaviour of the wind direction, of the temperature and of the sub-mesoscale to mean wind velocity ratio during events, in the most stable regimes. Events in Regimes S3 have a very large wind direction variability in SnoHATS and a slightly larger variability in F3 when compared to F1 and F2 fro FLOSSII. In Regimes S4 and F4, both characterised by a median $Ri$ value larger than $0.6$, the datasets differ markedly. The events of FLOSSII have a very large directional variability and large directional shifts. Such directional shifts were in fact shown to be common in the SBL under low wind speed by \cite{Mahrt:2010gv}, based on the FLOSSII data. In the latter study, the strongest wind-direction shifts were shown to occur often with a sharp decrease of temperature (a cold microfront). This was also found by \cite{Lang:2017bu} over a flat site in Australia. Moreover, \cite{Mahrt:2012fs} attributed an observed increase of wind directional shear at the FLOSSII site for increasing stratification to advection of cold air flow due to a cold pool forming upwind of the site. This is consistent with the events statistics for Regime F4, where the events are characterised by the largest wind directional shifts as well as the largest temperature changes, which are mostly negative (panels c, d and h). They are also characterised by the largest $\hat{V}/V$ ratios, which together with the wind directional variability could be a sign of wave-like activity. The events in the SnoHATS dataset however behave differently. Events in regime S3 shows signs of wave-like or advection activity similar to those seen in F4, with correspondingly larger ratio of $\hat{V}/V$. The temperature changes are larger in Regime S4 than in Regime S3, however Regime S4 has the least wind direction variability. In fact analysis of the wind direction distribution during events in Regime S4 point to a preferred direction pointing straight towards the instruments. This direction corresponds to an opening at the end of the glacier, forming a funnel that probably induces a wind direction constrained by the topography (Fig. \ref{fig:SnoHATS_setUp}). The temperature changes in this regime are for a majority cold temperature changes (cold microfronts) as highlighted by the statistics in Fig. \ref{fig:Physical_chara} panel (h). The cold microfronts events correspond to very little vertical velocity variance, and to rather small ratio of $\hat{V}/V$. This could indicate a regime where decreasing turbulence, surface cooling and increasing stratification could evolve together. The flow is near-calm and wave-like activity is no longer present. A similar regime is not detected in the FLOSSII dataset.

Regime S4 is thus characterised by submeso motions on scales significantly larger than the turbulent scales (as is better highlighted by analyses in \cite{Vercauteren:2016kx}), that take a slow microfront signature with little wind direction variability. We hypothesise that advected air masses or density currents that tend to take a microfront structure, while enhancing shear locally, may only trigger little turbulence on small scales. Regime F4 has a similar scale signature and microfronts structures, but the site features are such that the microfronts also correspond to large shifts in the wind direction. Nevertheless, these events also trigger only little turbulent mixing. On the contrary, the wind-direction variability characteristics of events in regime S3, with its scale overlap that was identified in \cite{Vercauteren:2016kx}, lead us to hypothesise that this regime encompasses wave-like phenomena that may break down to turbulence through a
cascade of scales. The velocity of the submeso motions in this regime is often larger than the wind velocity.

\subsection{Example of events and flow structures}
\label{subsec:Visualization}

\begin{figure}[!h]
	\centering
	\begin{subfigure}{0.58\textwidth}
		\includegraphics[width=\textwidth]{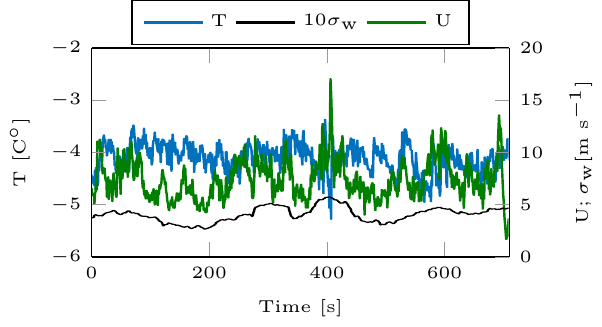}
		\caption{S1 event}
		\label{sfig:e1}
	\end{subfigure}
	\begin{subfigure}{0.38\textwidth}
		\centering
		\includegraphics[width=\textwidth]{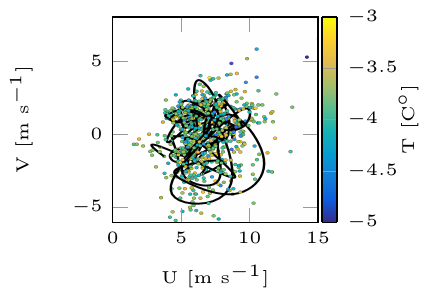}
		\caption{S1 phase space }
		\label{sfig:e1p}
	\end{subfigure}
	\quad
	\begin{subfigure}{0.58\textwidth}
		\includegraphics[width=\textwidth]{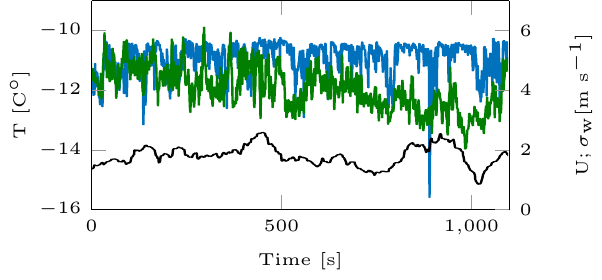}
		\caption{S2 event}
		\label{sfig:e2}
	\end{subfigure}
	\begin{subfigure}{0.38\textwidth}
		\centering
		\includegraphics[width=\textwidth]{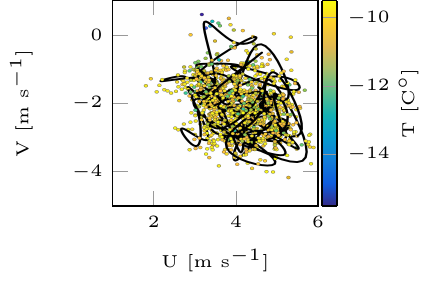}
		\caption{S2 phase space }
		\label{sfig:e2p}
	\end{subfigure}
	\quad
	\begin{subfigure}{0.58\textwidth}
		\includegraphics[width=\textwidth]{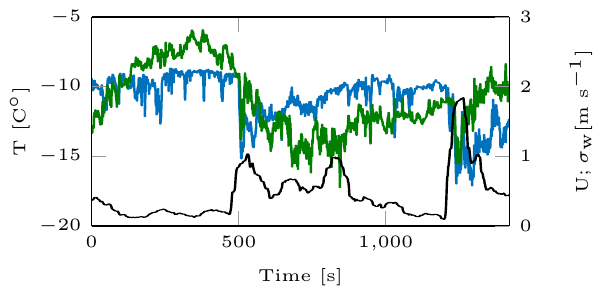}
		\caption{S3 event}
		\label{sfig:e3}
	\end{subfigure}
	\begin{subfigure}{0.38\textwidth}
		\centering
		\includegraphics[width=\textwidth]{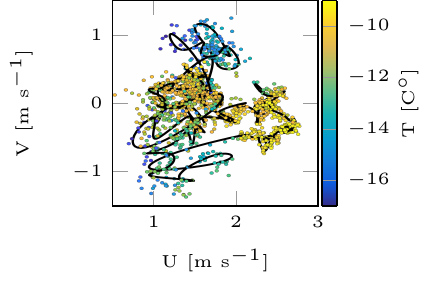}
		\caption{S3 phase space}
		\label{sfig:e3p}
	\end{subfigure}
	\quad
	\begin{subfigure}{0.58\textwidth}
		\includegraphics[width=\textwidth]{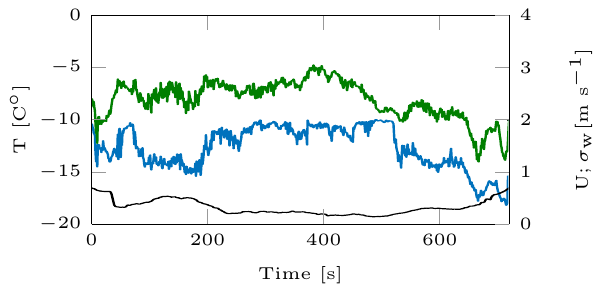}
		\caption{S4 event}
		\label{sfig:e4}
	\end{subfigure}
	\begin{subfigure}{0.38\textwidth}
		\centering
		\includegraphics[width=\textwidth]{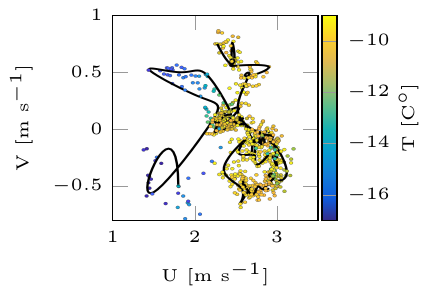}
		\caption{S4 phase space}
		\label{sfig:e4p}
	\end{subfigure}
	\caption{ Visualisation of turbulent events for the SnoHATS dataset. One 
		event is shown as an example for each regime.  The 
	timeseries on the left show the 
	temperature T (blue), the wind speed U (green) and the vertical velocity 
	component w (black), all shown for 6s averaged data. The scatterplots show the temperature in colour, in 
	the phase space of the horizontal velocity components. The black line is a 
	spline smoothing of the time evolution of the wind vector. 
	}\label{fig:SnoHATS_event3D}
\end{figure}

\begin{figure}[!h]
	\centering
	\begin{subfigure}{0.58\textwidth}
		\includegraphics[width=\textwidth]{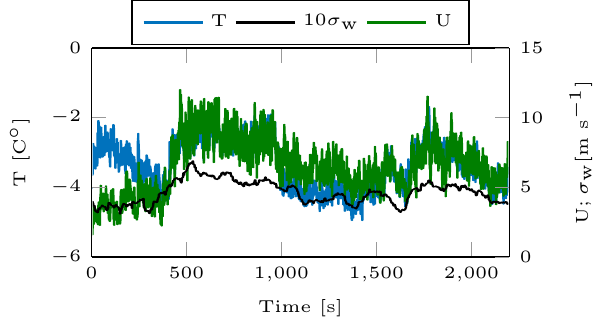}
		\caption{F1 event}
		\label{sfig:F_e1}
	\end{subfigure}
	\begin{subfigure}{0.38\textwidth}
		\centering
		\includegraphics[width=\textwidth]{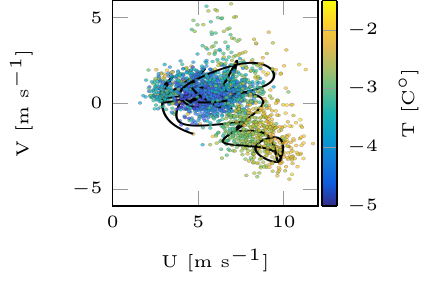}
		\caption{F1 phase space }
		\label{sfig:F_e1p}
	\end{subfigure}
	\quad
	\begin{subfigure}{0.58\textwidth}
		\includegraphics[width=\textwidth]{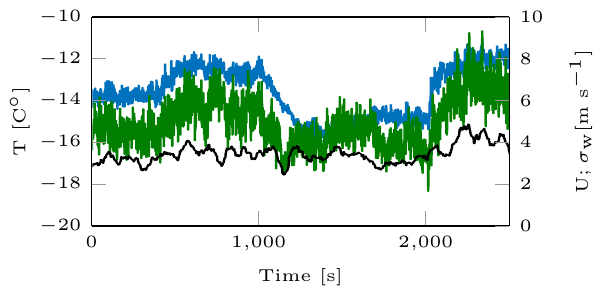}
		\caption{F2 event}
		\label{sfig:F_e2}
	\end{subfigure}
	\begin{subfigure}{0.38\textwidth}
		\centering
		\includegraphics[width=\textwidth]{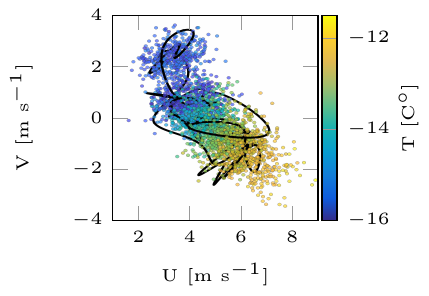}
		\caption{F2 phase space }
		\label{sfig:F_e2p}
	\end{subfigure}
	\quad
	\begin{subfigure}{0.58\textwidth}
		\includegraphics[width=\textwidth]{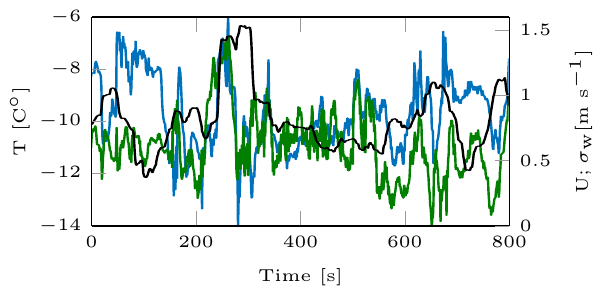}
		\caption{F3 event}
		\label{sfig:F_e3}
	\end{subfigure}
	\begin{subfigure}{0.38\textwidth}
		\centering
		\includegraphics[width=\textwidth]{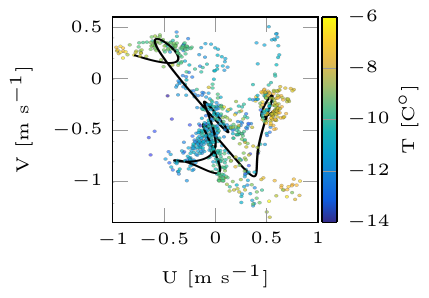}
		\caption{F3 phase space}
		\label{sfig:F_e3p}
	\end{subfigure}
		\quad
	\begin{subfigure}{0.58\textwidth}
		\includegraphics[width=\textwidth]{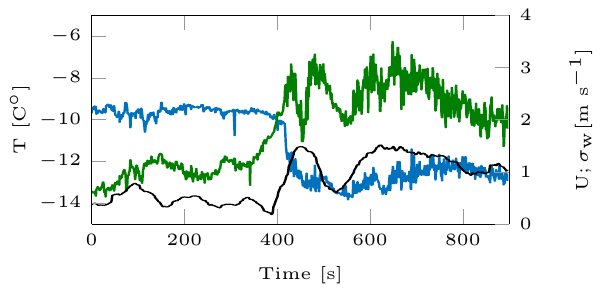}
		\caption{F4 event}
		\label{sfig:F_e4}
	\end{subfigure}
	\begin{subfigure}{0.38\textwidth}
		\centering
		\includegraphics[width=\textwidth]{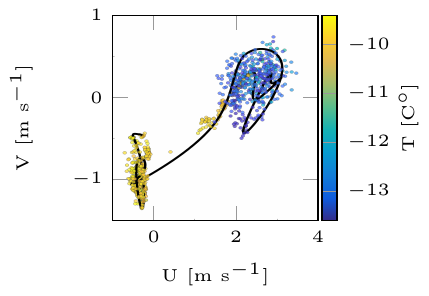}
		\caption{F4 phase space}
		\label{sfig:F_e4p}
	\end{subfigure}
	\caption{ Visualisation of turbulent events for the FLOSSII dataset. One 
		event extracted from the two meter measurement height is shown as an example for each regime.  The 
	timeseries on the left show the 
	temperature T (blue), the wind speed U (green) and the vertical velocity 
	component w (black), all shown for 6s averaged data. The scatterplots show the temperature in colour, in 
	the phase space of the horizontal velocity components. The black line is a 
	spline smoothing of the time evolution of the wind vector. 
	}\label{fig:FLOSSII_event3D}
\end{figure}

The events were detected by the TED method based solely on the temperature time series, without considering information on the wind direction. In this section we want to explore the flow structures corresponding to selected events, taken as example in each identified flow regime. One example of event is shown for each flow regime of SnoHATS in Fig. \ref{fig:SnoHATS_event3D}, and for each flow regime of FLOSSII in Fig. \ref{fig:FLOSSII_event3D}. Examples are chosen that approximately match the median characteristics illustrated in Fig. \ref{fig:Physical_chara}. The selection was done after visual analysis of all events, based on what appeared to be the most frequent or typical examples (additionally to matching the median characteristics approximately). In the examples selected from S1 and S2, the timeseries of temperature $T$ and wind speed $U$ show an oscillating behaviour which is not followed by $\sigma_w$ (Fig \ref{sfig:e1} and Fig \ref{sfig:e2}), as was observed by the FEM-BV-VARX clustering results (see Table \ref{tab:causality}). The corresponding evolution of the horizontal wind components during the events is shown in Fig \ref{sfig:e1p} and Fig \ref{sfig:e2p}, where the corresponding evolution of the temperature is represented by the colours of the scatterplot. The black line in the scatterplots is a smoothing of the time evolution of the wind vector, so as to smooth out the turbulence variability. This visualisation of the event highlights rather a mixing process, with mixed temperature (Fig \ref{sfig:e1p} and Fig \ref{sfig:e2p}). The timeseries of temperature $T$, wind speed $U$ and $\sigma_w$ for the events in F1 (Fig. \ref{sfig:F_e1}) and F2 (Fig. \ref{sfig:F_e2}) show that $T$ and $U$ tend to evolve in phase with a wavy pattern, but that $\sigma_w$ does not follow the dynamics of $T$ and $U$. This is in agreement with what was observed from the FEM-BV-VARX clustering analysis where no relationship was found between $\sigma_w$ and the sub-mesoscale wind variability (see Table \ref{tab:causality}). In Fig. \ref{sfig:F_e1p} and Fig. \ref{sfig:F_e2p}, the wind vector evolves in a compact structure, and the temperature changes smoothly following the wind vector. In these weakly stable regimes, events are present but $\sigma_w$ remains rather stationary during the events. The events of FLOSSII are more structured than at SnoHATS, which could be related to the differences in terrain complexity as discussed above.

In subsection \ref{subsec:FEM_BV_VARX-regime} we pointed out that sharp temperature changes in time occur in the most stable flow regimes S3-S4 and F4 (see Fig. \ref{sfig:Physical_chara_T}), being strongest in S4 and F4, while wind direction changes have a more site specific signature (see Fig. \ref{sfig:Physical_chara_Dir}). The wind direction variability during events is largest in S3, F3 and F4 and is visible in the examples from F3 and F4, both highlighting a dispatched structure in the time evolution of the wind vector (Fig. \ref{sfig:F_e3p}). The example of F4 has a sharp change of wind direction which is simultaneous to the sharp change of temperature (\ref{sfig:F_e4p}). This is the typical signature of a microfront which is commonly found with weak winds and thin stable boundary layers \citep{Mahrt:2010gv, Lang:2017bu}. For both examples, the dynamical evolution of $\sigma_w$ is highly non-stationary, partly affected by the evolution of $T$ and $U$ (Fig. \ref{sfig:F_e3} and Fig. \ref{sfig:F_e4}). In the example from S3, the time evolution of the wind direction seen in the phase space figure has a clear structure, but denotes less sharp transitions in wind direction than in the FLOSSII examples (Fig \ref{sfig:e3p}). The drop in temperature occurs sharply, with only a slight change in wind direction (visible in phase space as the transition from the homogeneous temperature part of the scatterplot on the right side to the left side of the scatterplot). Increases in $\sigma_w$ occur simultaneously to drops in the temperature (Fig. \ref{sfig:e3}), and the evolution of $T$ and $U$ is approximately in phase. In the example of S4 however, $\sigma_w$ seems rather stationary again, and the drop in temperature and wind speed does not correspond to a marked increase of $\sigma_w$ (Fig. \ref{sfig:e4}). The wind direction is oscillating (Fig. \ref{sfig:e4p}). This example could correspond to the radiative regime suggested in \cite{vandeWiel:2003uw}, with wind meandering but a collapsed state of turbulence. 
 
These examples are just a few of many events for which no clear dominant patterns were apparent. Regularity of events signatures was searched for using the clustering method suggested in \cite{Kang:2015hi}, which highlighted reoccurring events signatures in the FLOSSII dataset. For the SnoHATS dataset however, similar regularity was not made apparent through that clustering approach, and did not appear after visual inspection either. The classes of sub-mesoscale and turbulence interaction analysed in our study identify different regimes which may have a preference for the existence or absence of microfronts of temperature, sharp changes in wind direction or wind oscillations. However the specific types of submeso motions occurring may be more dependent on the local terrain features and less on the flow type. We expect flow regimes with active turbulence to quickly dissipate submeso motions, whereas quieter regimes allow such flow structures to be transported for a longer time and thereby to affect the turbulence to some extent. This picture is consistent with the hypothesis of a dynamically unstable flow in a transition zone near a critical wind speed separating two distinct metastable flow equilibria (one with active turbulence and one with collapsed turbulence), suggested by the conceptual model of \cite{vandeWiel:2017ju}. Indeed, in the transition zone near the critical wind speed, the response time of the flow to perturbations becomes large, such that random perturbations of the flow by submeso motions are long-lived. This is in agreement with the statistical analyses presented in \cite{Nevo:2017ir} that suggested that the more stable regimes are dynamically unstable, based on analysing memory effects in timeseries of turbulent observables in the different flow regimes.

\section{Conclusion}
\label{sec:conclusion}
Flow regimes were classified in terms of interactions between submeso and turbulent scales of motion. In each flow regime, turbulent events were extracted using the TED method, and the statistical properties of those events were characterised. Regimes experiencing little scale interactions (S1, S2, F1, F2) are characterised by the shortest events, higher wind speeds, weak stability and fewer events.

Regimes experiencing more scale interactions correspond to higher stability, more numerous and longer events. In the most stable regimes that occur with weak winds, with a scale separation between turbulent and sub-mesoscales, the signature of events was found to take a site specific signature, probably related to local topographical characteristics. Events in these very weak wind conditions tend to exhibit strong temperature changes, with wind direction variability characteristics that depend on the site, probably through the terrain specificities. The site differences exemplified here on two datasets render the derivation of parameterisations difficult. These flow regimes could be related to radiative cooling, advected air masses, density currents, and events thus tend to take a microfront structure with sharp temperature changes. Local shear enhancement due to the advected air masses results in turbulent mixing as identified by the FEM-BV-VARX method but the turbulent mixing occurs on very local, small scales. There may be a direct energy transfer between sub-mesoscales and turbulent scales through local shear generation of small-scale eddies, as was analysed in \cite{Vercauteren:2016kx}.

In regimes where the sub-mesoscale wind velocity is often larger than the wind velocity (S3, F3), or where the submeso and turbulent scales tend to overlap (S3), events are characterised by a large variability in the wind direction during the selected event. Temperature changes are however less sharp than in the most stable regimes S4 or F4 in which turbulence is nearly collapsed. Events are associated with stronger vertical velocity fluctuations than in the very weak wind, strongly stable regimes S4 and F4. This could potentially be more related to wave-like phenomena that break down to turbulence through a cascade of scales.

The MRD analyses point to the randomness of sub-mesoscale contributions. The averaged contribution of the sub-mesoscales to the heat flux is negligible, however individual contributions become larger than turbulent contributions in strongly stable, weak wind regimes. The phenomena leading to the sub-mesoscale motions and associated fluxes are not resolved nor taken into account in numerical models, except through artificial enhanced mixing. The combination of flow regime detection to extract periods where submeso motions tend to dominate over turbulent transport and characterisation of events in each regime provides a way to define a stochastic process based on the statistical analyses. More extensive analyses based on machine learning methods, for example using neural network methods to identify patterns using all the observables corresponding to events may help shed light on the regularity of submeso motions.

\section*{acknowledgements}
The authors thank Marc Parlange and the EFLUM laboratory at EPFL for providing the SnoHATS data and Larry Mahrt for providing the FLOSSII data and help to analyse them. Illia Horenko and Dimitri Igdalov provided the FEM-VARX framework and help that was greatly appreciated. Alexander Kuhn is greatly acknowledged for his software provided to visualise flow structures. The research was funded by the Deutsche Forschungsgemeinschaft (DFG) through grant number VE 933/2- 1 "Towards Stochastic Modelling of Turbulence in the Stable Atmospheric Boundary Layer", and has been partially supported by DFG through grant CRC 1114 "Scaling Cascades in Complex Systems", Project (B07) "Selfsimilar structures in turbulent flows and the construction of LES closures". 

\clearpage
\pagebreak
%
\bibliographystyle{spbasic_updated}      
\bibliography{bibliography}   
\end{document}